**5**

# PHOTOMETRIC STUDY OF THE SHORT PERIOD W UMA SYSTEM VW CEPHEI

**Ahmed-Essam, A. E. Alawy and A. A. Haroon**

**ABSTRACT:** *A total of 431 photoelectric B and V observations are presented for the short period W UMa system VW Cep. The measurements have been obtained in two successive nights, 27/28 and 28/29 of September 1995, and represent the whole light curve phases. Four times of minimum light, for each filter, were deduced and new periods have been derived as 0.277892 (±0.00013) day from the present photometric observations. New light curves have been constructed, investigated and analyzed by using the recent version of W-D code.*

## INTRODUCTION

The short period variable star VW Cephei (HD 197433; BD +75° 752; SAO 9828) is fairly available for observing continuously, because the system exhibits several phenomena that are characteristics of W-type group of the W UMa eclipsing binaries. Some of these are:
1. Short period (less than 7 hours) accompanied by period variations.
2. The light variation exhibits pronounced O'Connell-effect.
3. The system components show variable surface activity changing the shape of the light curve and the depth of eclipses.

## PRESENT OBSERVATION

The orbital period of VW Cep is short, P=0.277892 day, therefore it is possible to observe the whole orbital cycle during one night. Photoelectric observations were carried out through B and V wide passband filters in 2 nights, 27/28 and 28/29 of September 1995, using the Kottamia 74" Telescope. The stars BD=75° 726 (HD 192889; SAO 9669) and BD=73° 900 (HD 192635; SAO 9659) have been used as a comparison and checking, respectively. Table (1) represents the observational data of the system VW Cep and the reference stars.

**Table 1: Observational Data of the System VW Cep.**

| Star Name | BD No. | R. A. 2000 h   m   s | Dec. 2000 °   ′   ″ | V. Mag. | Spec. Type | No. of obs. B | V |
|---|---|---|---|---|---|---|---|
| VW Cep. | 75° 752 | 20  37  21.5 | 75  36  01.5 | 7.2-7.7 | K0V | 431 | 431 |
| Comp. | 75° 726 | 20  09  40.0 | 76  14  41.9 | 8.2 | G5 | 21 | 21 |
| Check | 73° 900 | 20  09  15.3 | 74  25  32.3 | 8.9 | F4IV | 21 | 21 |

A total of 431 observations for each B and V filters were achieved. Two light curves for each of the primary and secondary eclipses were obtained. The heliocentric Julian date (HJD), and the differential magnitude (Delta Mag.), in the sense of variable minus comparison, were calculated and plotted in Figures (1) and (2). From these Figures, we can notice that, the heights of the two maxima are not equal (Max.II - Max.I = 0.038 mag.)$_V$, (Max.II - Max.I = 0.0285 mag.)$_B$. The two minima are of unequal depth (Min.II - Min.I = 0.078 mag.)$_V$, (Min.II - Min.I = 0.088 mag.)$_B$. A shoulder is seen in Figure (1) at the values 8.31, 8.36 of HJD, mostly in the B band. Other shoulders appear in Figure (2) at the values 9.28, 9.30, and 9.32 of HJD in the B band, but it is not detected in the V band. We can notice an unusual large scatter in both B and V bands around the primary maximum. Also, we noticed a quite night-to-night variation similar to the finding of Rovithis et al. (1980).

## EPOCHS OF PHOTOMETRIC MINIMA

The times of minima for the system VW Cep were calculated using the software package AVE (Barbera, 1996), which employs the method of Kwee-Woerden (1956). The deduced times of minima are listed in Table (2) together with their probable errors (P.E.), type of minimum (Min), and the band of observations used (Filter).

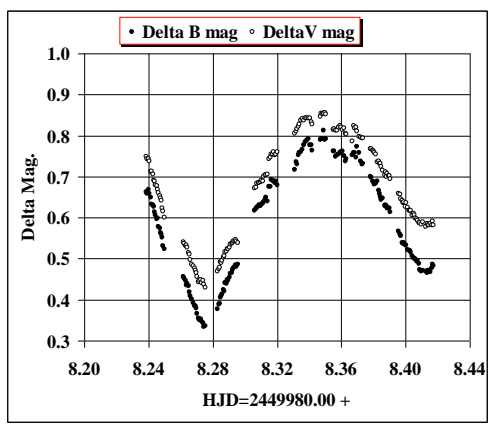

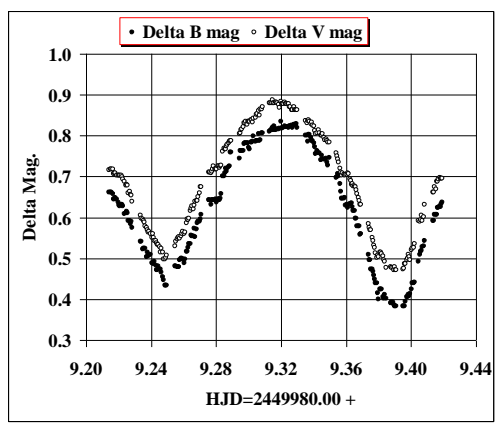

**Fig. 1: Delta B and V Observations of VW Cep on 27/28 Sep. 1995**

**Fig. 2: Delta B and V Observations of VW Cep on 28/29 Sep. 1995**

_______________________________________________________________

Table 2: Epochs Minimum Light of VW Cep.

| Date | HJD=2449900+ | P.E. | Min | Filter |
|------|-------------|------|-----|--------|
| 27 September 1995 | 88.275910 | ± 0.000149 | I | B |
| | 88.276124 | ± 0.000159 | I | V |
| | 88.411809 | ± 0.000113 | II | B |
| | 88.412678 | ± 0.000279 | II | V |
| 28 September 1995 | 89.390613 | ± 0.000298 | I | B |
| | 89.390277 | ± 0.000186 | I | V |
| | 89.249502 | ± 0.000174 | II | B |
| | 89.249176 | ± 0.000268 | II | V |

## PHOTOMETRIC PERIOD DETERMINATION

The photometric period of the system has been derived via AVE package (Barbera, 1996), where Fourier series have been applied. The observations achieved on 27 and 28 September 1995 have been combined. This is because of two reasons, first; both minima and maxima were observed; i.e. all characteristic phases are well represented. Secondly, the observations have been obtained in short time span (two consecutive nights). This may minimize errors resulted from long-term light curve variation.

A period of 0.277892 (±0.000134) days, as the mean of those derived from blue and visual observations, has been considered. Adopting the mean time of primary minima of the first night (27 September 1995) and the new period, we have obtained the new ephemeris which represent well the present data as follows:

Hel. J.D. (Min.I) = 2449988.276017 (± 0.000151) + 0.277892 (± 0.000134) * E

The observations have been phased using this ephemeris and transformed into standard B and V magnitudes. The normal points of the light curves have been derived, as seen in Figure (3), and were employed for light curve modeling.

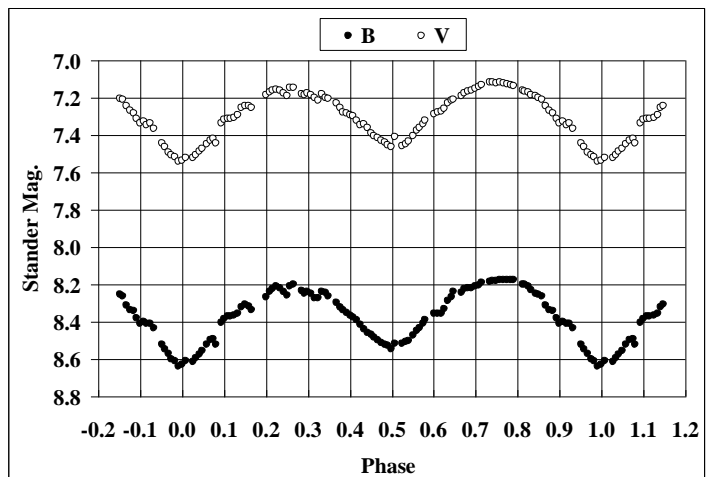

**Fig. 3: Normal points B and V Standard Mag. of VW Cep**

# GEOMETRICAL AND PHYSICAL PARAMETERS

Applying cosine Fourier series technique (see Rucinski, 1993) leads to the coefficients $A_1$ to $A_{10}$ of the V, and B light curves. Using the three coefficients ($A_2$, $A_4$, and $A_6$) with the tables of Rucinski, we can estimate the three main geometrical parameters, inclination (i), mass ratio (q), and degree of contact (f), as follows: i ~ 67,     q ~ 0.25,   and    f ~ 0.5
The values of i, q, and f, have been used as initial and control values for the light curve analysis.

The normal points of the standard B and V light curves in Fig. 3 have been used for the present analysis using the recent version of Wilson-Devinney programme (W-D, 2003). The model has been described and quantified in papers by Wilson & Devinney (1971), Wilson (1979, 1990, 1993), and Van Hamme & Wilson (2003).
Most of the previous studies (e.g. Kaszas et al. 1998) show that the system is contact and belongs to the W-Type of W UMa systems. So, mode 3 was adopted to carry out the photometric solution. Using the first part of W-D program (Light curve programme; LC), we tried with different sets of parameters through V and B Filters to find a reasonable solution. Starting without any spots on both stars, no reasonable fit was

_______________________________________________________________________

obtained. Therefore, we decided to adopt only one spot on the cool star (second component). After many trials, through changing some parameters (inclination, surface temperature of both stars and surface potential) we got a fine fit curves for V and B filters. Applying the second part of W-D programme (Differential Correction programme; DC), with the initial parameters values obtained from the last run of LC Program. The temperature of hot star ($T_1$) was adjusted while $T_2$ was kept fixed. The phase shift, orbital inclination, surface potential, mass ratio, and spot temperature were adjusted. The final solutions have been plotted in Figures (4) and (5) for V and B filter respectively whose parameters are tabulated in Table (3) with their standard deviation.

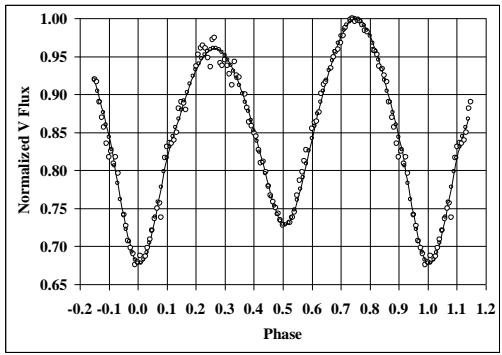

**Fig. 4: Fit curve with the normalized flux for all V observations of VW Cep**

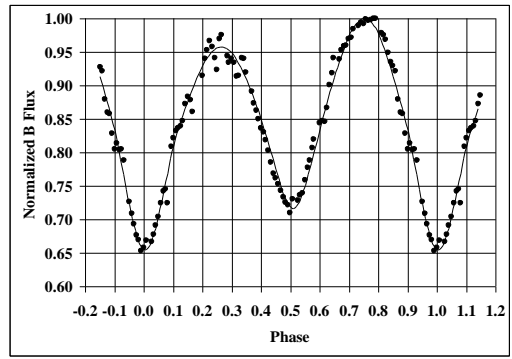

**Fig. 5: Fit curve with the normalized flux f or all B observations of VW Cep**

The fit is good enough in the V and B light curves except near the primary maximum in both filters, where there is a deviation between the computed values and the observations. This might be due to magnetic activities of cool star. The geometrical and physical parameters and the Binary Maker 3.0 program (Bradstreet 2004) have been used to produce the Roche geometry with the degree of contact as shown in Figure (6). Also system configuration at different phases with the spot adopted is displayed in Figure (7).

Bradstreet (2004) have been used to produce the Roche geometry with the degree of contact as shown in Figure (6). Also system configuration at different phases with the spot adopted is displayed in

Figure (7).

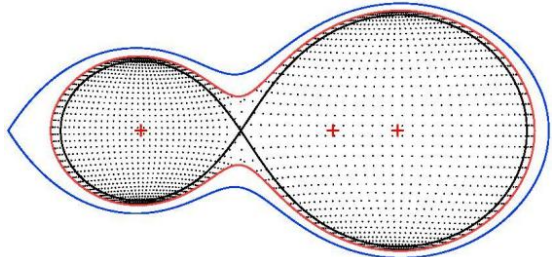

**Fig. 6: Roche Geometry of the System VW Cep.**

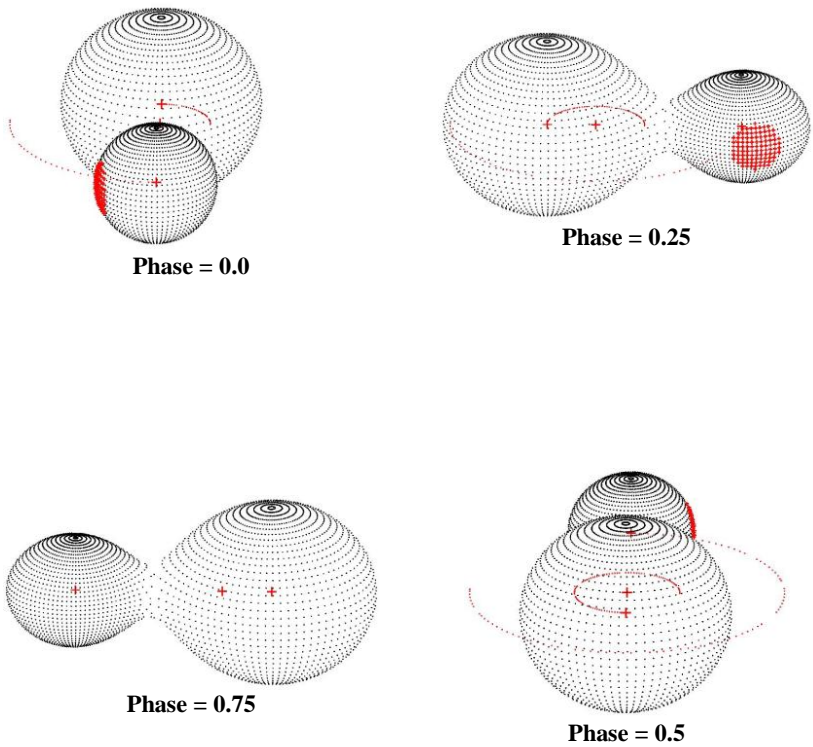

**Fig. 7: Shape of the system VW Cep at different phases**

**Table 3: Adjusted Parameters of VW Cep, for V and B filters**

| Parameter | V filter | B filter |
|---|---|---|
| Phase Shift | $00.0033 \pm 0.0009$ | $00.0022 \pm 0.0009$ |
| Inclination (i) | $67^{°}.418 \pm 0^{°}.3685$ | $68^{°}.625 \pm 0^{°}.5647$ |
| Semi major axis | 1.8577 (fixed) | 1.8577 (fixed) |
| Mass Ratio ($q=m_2/m_1$) | $0.3349 \pm 0.0040$ | $0.3627 \pm 0.0085$ |
| Surface Temp. $T_1$ | 4970 k $\pm$ 13.58 k | 4949.5 k $\pm$ 32.65 k |
| Surface Temp. $T_2$ | 4650 k (fixed) | 4700 k (fixed) |
| Surface potential ($\Omega_1 = \Omega_2$) | $2.4588 \pm 0.0098$ | $2.5406 \pm 0.0216$ |
| $\Omega_{Inner}$ | 2.5409 | 2.5423 |
| $\Omega_{Outer}$ | 2.3339 | 2.3350 |
| Bolometric albedo ($A_1=A_2$) | 0.500 (fixed) | 0.500 (fixed) |
| Gravity exponents ($\alpha_1= \alpha_2$) | 0.320 (fixed) | 0.320 (fixed) |
| Third light | 0.0100 (fixed) | 0.0100 (fixed) |
| $L_1 / (L_1+ L_2)$ | $0.8060 \pm 0.0297$ | $0.7970 \pm 0.1050$ |
| $L_2 / (L_1+ L_2)$ | 0.2040 (calculated) | 0.2030 (calculated) |
| Limb Darkening ($x_1$) | 0.726 (fixed) | 0.867 (fixed) |
| Limb Darkening ($x_2$) | 0.779 (fixed) | 0.914 (fixed) |
| Latitude of spot No. 1 | $90^{°}.00$ (fixed) | $90^{°}.000$ (fixed) |
| Longitude of spot No. 1 | $105^{°}.00$ (fixed) | $105^{°}.00$ (fixed) |
| Radius of spot No. 1 | $25^{°}.00$ (fixed) | $27^{°}.000$ (fixed) |
| ($T_{spot} / T_{star}$) of spot No.1 | $0.7589 \pm 0.1341$ | $0.8734 \pm 0.1782$ |
| % Overcontact | 36 % | 42 % |
| $r_1$(pole) | $0.4649 \pm 0.0022$ | $0.4649 \pm 0.0048$ |
| $r_2$(pole) | $0.2884 \pm 0.0035$ | $0.2884 \pm 0.0076$ |
| $r_1$(side) | $0.5034 \pm 0.0031$ | $0.5034 \pm 0.0068$ |
| $r_2$(side) | $0.3034 \pm 0.0044$ | $0.3034 \pm 0.0095$ |
| $r_1$(back) | $0.5360 \pm 0.0043$ | $0.5360 \pm 0.0095$ |
| $r_2$(back) | $0.3540 \pm 0.0097$ | $0.3540 \pm 0.0208$ |
| $M_1$ & $M_2$ (Masses in Solar Units) | 0.838 & 0.280 | 0.820 & 0.297 |
| $R_1$ & $R_2$ (Radius in Solar Units) | 0.93 & 0.58 | 0.91 & 0.58 |
| $M_{bol}$ & $M_{bol}$ | 5.62 & 6.91 | 5.67 & 6.86 |
| Log $g_1$ & Log $g_2$ | 4.43 & 4.36 | 4.44 & 4.38 |
| $\Sigma \omega$ (O-C)$^2$ | 0.0153 | 0.01336 |

The fit is good enough in the V and B light curves except near the primary maximum in both filters, where there is a deviation between the computed values and the observations. This might be due to magnetic activities of cool star. The geometrical and physical parameters and the Binary Maker 3.0 program (Bradstreet 2004) have been used to produce the Roche geometry with the degree of contact as shown in Figure (6). Also system configuration at different phases with the spot adopted is displayed in Figure (7).

_______________________________________________________________________

# DISCUSSION AND CONCLUSION

The B and V photometry presented here has been obtained in two consecutive nights covering all phases of the light curves and including two primary and two secondary minima. Light variation of VW Cep, exhibits the famous phenomena of W UMa systems, like:

a- O'Connell effect variation.

(Max. II - Max. I = 0.038 mag.)$_V$,  (Max. II - Max. I = 0.0285 mag.)$_B$.

b- The two minima are of unequal depth.

(Min. II - Min. I = 0.078 mag.)$_V$,  (Min. II - Min. I = 0.088 mag.)$_B$.

c- Asymmetric minima concerning the ingress and egress eclipses' branches. Recent spectroscopic study by Frasca et al. (1996) showed that the cooler component posses variable $H_\alpha$ emission correlated with the system photospheric spots. As a consequence of their study, it has been suggested that the light curve asymmetry is caused by the mass flow from the cooler component to its companion through the inner Lagrangian point.

d- A shoulders were observed at different phases, mostly in the B band similar to that noticed by Rovithis, et al. (1980).

e- The two maxima exhibit pronounced distortion as small humps and short duration dips. These have been explained by Vilhu and Heise (1998) and Vilhu et al. (1988), from their X-ray and radio 6 cm observations, by suggesting absorbing clouds above small emitting regions close to the system neck. These authors detected flares similar to the solar ones and showed that VW Cep behaves as the Sun before, during, and after flare occurrence.

f- From practical point of view, the scatter presented in the light curve is a permanent feature of VW Cep. A value of about 0.15 mag. has been found by Bradsteet and Guinan (1990) at all phases for visual observations achieved over about 12 years.

g- Times of two primary and two secondary minima have been determined from our photometry. A new linear ephemeris was determined as:

Hel.J.D. (Min.I) = 2449988.276017 (±0.000151) + 0.277892 (±0.000134) * E

h- Three plausible sets of solution of physical and geometrical parameters have been derived, from the present light curves analyses, showing different spots on the system hot and cool components as follows:

_______________________________________________________________________

|            | H. Star (Smaller) | C. Star (Larger) | $\Sigma \omega (O-C)^2$ |
|------------|-------------------|------------------|-------------------------|
| Solu. (1)  | ----------        | 3 Spots          | 0.01102                 |
| Solu. (2)  | ----------        | 1 Spot           | 0.01267                 |
| Solu. (3)  | 3 Spots           | ----------       | 0.00850                 |

All solutions have almost the same $\Sigma \omega (O-C)^2$ values. However, from the 3-D plots of the system in the first and third solutions, we found that the angular radius of some spots is larger compared with the size of sunspots, that is may not be acceptable. The final and reasonable acceptable solution is the solution (2) listed in Table (3) and shown in Figures 4, 5, 6, and 7.

From the light curves' solutions, we may state that the spot model cannot provide unique photometric solution.

## ACKNOWLEDGEMENT

We are grateful to Prof. Dr. R. Wilson for his data analyses programme (W-D code) and Prof. Dr. D. Bradstreet for his package Binary Maker 3.0, which we used as a pre-analyses programme and to illustrate the 3-D configuration of the system at different phases.